# On-the-fly particle metrology in hollow-core photonic crystal fibre

Abhinav Sharma[1], Shangran Xie[1]*, Richard Zeltner[1], and Philip St.J. Russell[1,2]

[1]Max Planck Institute for the Science of Light and [2]Department of Physics, Friedrich-Alexander University, Staudtstr. 2, 91058 Erlangen, Germany

*shangran.xie@mpl.mpg.de

Efficient monitoring of airborne particulate matter (PM), especially particles with aerodynamic diameter less than 2.5 μm ($PM_{2.5}$), is crucial for improving public health[1,2]. Reliable information on the concentration, size distribution and chemical characteristics of PMs is key to evaluating air pollution and identifying its sources[3]. Standard methods for $PM_{2.5}$ characterization require sample collection from the atmosphere and post-analysis using sophisticated equipment in a laboratory environment, and are normally very time-consuming[4,5]. Although optical methods based on analysis of scattering of free-space laser beams or evanescent fields are in principle suitable for real-time particle counting and sizing[6-9], lack of knowledge of the refractive index in these methods not only leads to inevitable sizing ambiguity but also prevents identification of the particle material. In the case of evanescent wave detection, the system lifetime is strongly limited by adhesion of particles to the surfaces[8,9]. Here we report a novel technique for airborne particle metrology based on hollow-core photonic crystal fibre. It offers in situ particle counting, sizing and refractive index measurement with effectively unlimited device lifetime, and relies on optical forces that automatically capture airborne particles in front of the hollow core and propel them into the fibre. The resulting transmission drop, together with the time-of-flight of the particles passing through the fibre, provide unambiguous mapping of particle size and refractive index with high accuracy. The technique represents a considerable advance over currently available real-time particle metrology systems, and can be directly applied to monitoring air pollution in the open atmosphere as well as precise particle characterization in a local environment such as a closed room or a reaction vessel.

Environmental air pollution, caused by mixtures of solid and liquid microparticles, is of growing public concern for public health[10-13]. Particles of diameter less than 2.5 μm ($PM_{2.5}$) are especially hazardous since they can penetrate deep into the respiratory system, causing severe damage[2-4]. A variety of non-optical techniques for monitoring $PM_{2.5}$ particles have been developed in recent years, involving atmospheric sampling through impingement on an adhesive surface or filtering, followed by analysis using bulky and sophisticated equipment[14-16]. They come with the disadvantage that sampling and inspection are not simultaneous, and that analysis is time-consuming and relatively high cost.

Optical methods based on detection of forward-scattered laser light offer a simple and inexpensive solution for real-time measurement of particle concentration, and form the basis of a number of commercial products[6,7]. Although particle sizing is also possible by detection of side-scattered light[17], environmental fluctuations cause random particle trajectories across the laser beam, making accurate particle sizing very difficult[18,19]. To overcome these limitations, the evanescent fields surrounding whispering-gallery-mode resonators on silicon chips[8] or glass nanofibres[9] have been used to size particles as small as 60 nm in diameter. In order to interact efficiently with the strongly confined evanescent field, however, the particles must adhere to the surface of the devices, requiring complicated cleaning procedures after each detection cycle[8,9] and unavoidably limiting device lifetime. In all these methods, the scattering cross-section of each particle depends not only on its diameter but also its (usually unknown) refractive index. This introduces uncertainty in particle sizing[20], precluding identification of the chemical composition of the particle, which is correlated with its refractive index.

Here we report a novel approach to real-time particle characterization that elegantly overcomes these limitations. It is based on hollow-core photonic crystal fibre (HC-PCF), and is capable of counting, sizing and measuring the refractive index *in situ* without the need to attach particles to a surface. The concept is illustrated in Fig. 1a. A focused laser beam is launched into the $LP_{01}$-like fundamental mode of the HC-PCF and a stream of airborne particles from a nebulizer (see Methods) is directed towards it from above. Single particles are randomly captured by the laser beam, held against gravity by optical gradient forces ($\mathbf{F}_g$), and propelled into the hollow core by optical scattering forces ($\mathbf{F}_s$). After a particle has entered the fibre, scattering causes the transmission (initially $\tau_0$) to fall to $\tau_p$ in the presence of a particle. Typical transmission drops $\delta\tau = (\tau_0 - \tau_p)$ recorded by the photodetector (PD) when polystyrene and silica particles of different diameters are propelled through the hollow core, are plotted in Figs. 1b and 1c. It is clear that $\delta\tau$ depends on both the particle diameter $d$ and its refractive index $n_p$, as does the time-of-flight $T_f$ of the particle through the fibre. We will show below that $\delta\tau$ and $T_f$ together can be used to calculate particle size and refractive index with high accuracy. After the particles are expelled from the fibre, the transmitted signal recovers to its initial value (Figs. 1b and 1c), with no sign of degradation even after many particles have travelled through the fibre. Optomechanical trapping of the particle within the protected environment of the hollow core leads to highly reproducible results, and ensures high accuracy and good repeatability in the measurements.

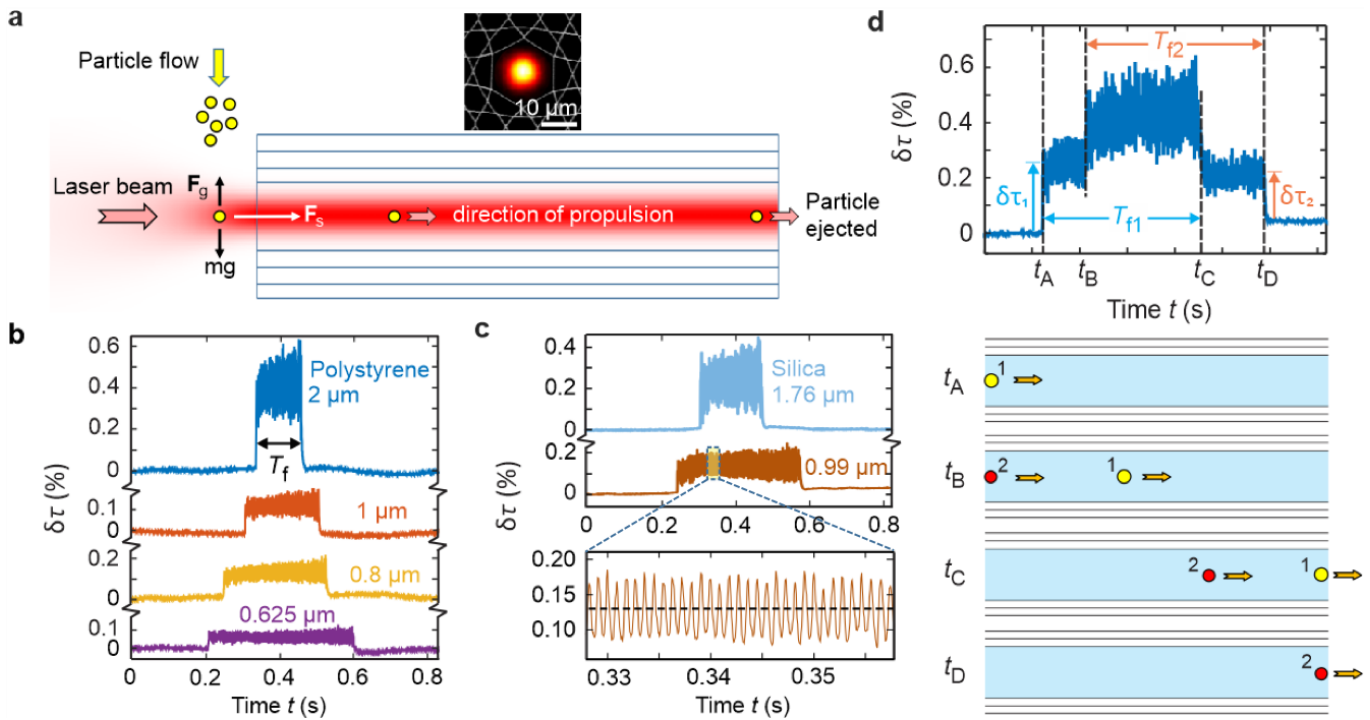

**Fig. 1 | Particle metrology in HC-PCF. a,** Sketch illustrating the operating principle of the particle detector. A focused laser beam captures an airborne particle in front of the fibre endface, where the optical gradient force ($F_g$) counter-balances gravity, and a strong scattering force ($F_s$) propels the particle into and along the hollow core. The transmission drops when the particle is in the core, and recovers after the particle has been expelled. Inset: measured near-field optical micrograph of the core mode superimposed on a scanning electron micrograph of the HC-PCF structure. **b,** Measured transmission drops $\delta\tau$ when polystyrene particles of different diameter were detected. **c,** Upper: transmission drop for silica particles. Lower: zoom-in of the yellow-shaded area in the upper panel, showing the effects of intermodal beating. The black-dashed line marks the average value. **d,** Transmission drop when polystyrene particles with 1.65 μm (yellow) and 1 μm diameter (red) were sequentially launched into the HC-PCF. The drop in transmission after the two particles have left the fibre is caused by a slow drift in the laser power.

In the experiment the transmission $\tau_p$ oscillates up and down with time, as seen in the lower panel in Fig. 1c, which is a magnified version of the yellow-shaded area in the upper plot (0.99 μm silica particle). This is caused by intermodal beating between the $LP_{01}$ and (accidentally excited) $LP_{11}$ core modes, which has the effect of driving the particle along a sinusoidal trajectory. We used the average value of $\tau_p$ (marked by the black-dashed line) to calculate $\delta\tau$. We found that this beat pattern also assisted in differentiating the particle-related signal from laser intensity noise.

The technique also works with a few particles, as seen in Figure 1d for the case of sequential loading of two polystyrene particles. At time $t_A$, a 1.65 μm particle (yellow) enters the fibre and the transmission drops to $\tau_{p1}$. At time $t_B$ a second particle (1 μm, red) enters the fibre before the first has left, and the transmission drops to $\tau_{p12}\tau_{p2}$, where $\tau_{p12}$ (the transmission by particle 1 after scattering by particle 2) is not in general equal to $\tau_{p1}$ because excitation of higher-order modes (HOMs) by the particle 1 affects scattering by particle 2. At $t_C$, the first particle exits the fibre and the transmission rises to $\tau_{p2}$, returning to $\tau_0$ after the second particle has left the fibre ($t_D$). The measurement also yields the times-of-flight of the particles. For the second particle $T_{f2} = L/[\alpha P_0(1-\tau_{p2})]$, where $L$ is fibre length, $P_0$ is the power impinging on the particle and $\alpha$ is a constant with units m.s$^{-1}$W$^{-1}$. For the first particle the expression is more complicated, requiring a value for $\tau_{p12}$ (not straightforward to derive).

The system offers reliable long-term monitoring, one at a time, of a sequence of single particles. A two-dimensional scatterplot of $\delta\tau$ and $T_f$, measured for polystyrene (dots) and silica (squares with black outline) particles with different diameters is shown in Figure 2a. The data-sets are also fitted to Gaussian distributions along the $\delta\tau$ and $T_f$ axes. The transmission drop increases and the time-of-flight falls as the particle diameter $d$ rises, as expected since the radiation pressure scales with ~$d^2$ while the viscous drag scales with ~$d$. Interestingly, smaller particles are more easily distinguished by time-of-flight, and larger particles by transmission drop, permitting a wide range of particle diameters to be clearly identified on this two-dimensional map.

The grey solid and dashed lines in Figure 2a, generated using the theory described below, map out contours of constant refractive index and particle size. Silica particles follow contour lines with refractive index of ~1.45, and polystyrene particles a refractive index of ~1.57, in agreement with the values specified by the particle manufacturer. All the particles tested are located approximately at the intersections between the appropriate contour lines of particle size and refractive index. The points located away from the main peaks on the histograms (marked with triangles) we attribute to particle clusters formed in the nebulizer (see Methods).

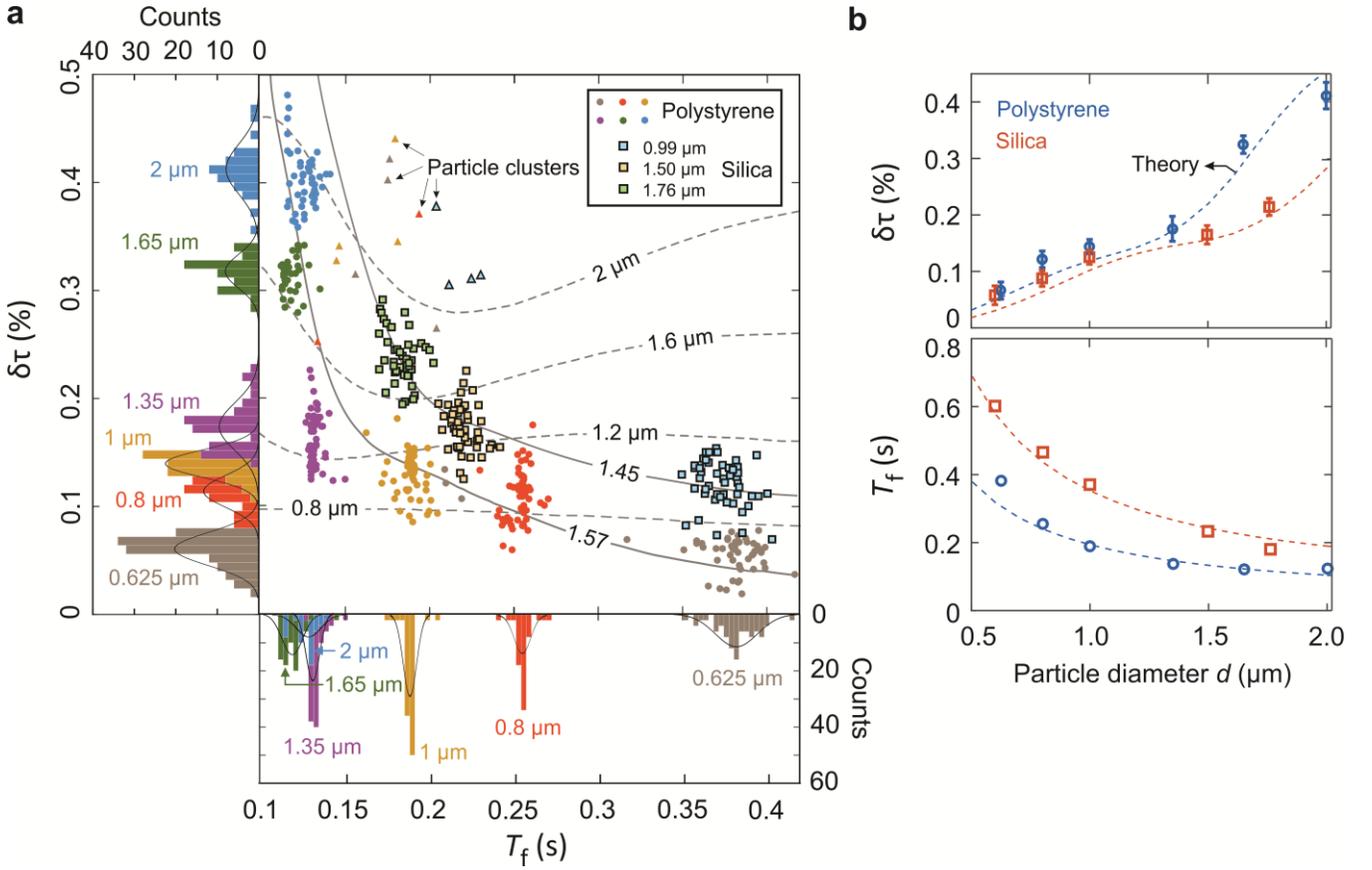

**Fig. 2 | Mapping of particle diameter $d$ and refractive index by transmission drop $\delta\tau$ and time-of-flight $T_f$. a,** Two-dimensional scatterplot of $\delta\tau$ and $T_f$ for polystyrene (dots) and silica (squares with black outlines) particles, together with histograms for polystyrene particles. The black lines in the histograms represent Gaussian fits. The solid grey lines are contours of constant refractive index and the dashed grey lines contours of constant $d$. We attribute the out-lying points, marked with triangles, to occasional particle clusters. **b,** $\delta\tau$ (upper) and and $T_f$ (lower) plotted against particle diameter $d$ for polystyrene (blue circles) and silica particles (orange squares). The error-bars mark the standard deviations of Gaussian fits to the histograms, and are smaller for $T_f$ than $\delta\tau$. The dashed curves are predictions from the theory described in the text, without any free parameters.

A particle optically trapped in the hollow core scatters the incident light in both backward and forward directions. The reflection coefficient, which directly introduces a drop in the fibre transmission and depends on the scattering cross-section between the particle and the $LP_{01}$-like core mode, can be written as:

$$\rho = \left(\frac{n_p-1}{n_p+1}\right)^2 \frac{\int_0^{d/2} J_0^2(2u_{01}r/D)r dr}{\int_0^{D/2} J_0^2(2u_{01}r/D)r dr} \quad (1)$$

where $n_p$ is the particle refractive index, $D$ the core diameter, $u_{01}$ the first zero of the Bessel function $J_0$ and the factor outside the integrals is the Fresnel reflection coefficient. In the forward direction the centrally-placed particle scatters a fraction of the incident $LP_{01}$ light into azimuthally invariant HOMs, with transmission coefficients given by[22]:

$$\tau_{0m} = \left|\frac{\iint e_s e_{0m}^* dA}{\iint |e_{01}|^2 dA}\right|^2 \quad (2)$$

where the integrations are over the transverse plane and $e_{01}$, $e_{0m}$ and $e_s$ are respectively the field distributions of the $LP_{01}$ mode, the $LP_{0m}$ mode and the forward-scattered field just after the particle. Since HOMs are more divergent than the $LP_{01}$ mode, they can be efficiently filtered out by placing an aperture in front of the PD detector. The total drop in transmission of the $LP_{01}$ mode is then $\delta\tau = 1 - \tau = (\rho + \Sigma\tau_{0m})$, where the summation is for $m \geq 2$. The upper part of Figure 2b plots the measured values of $\delta\tau$ obtained from Gaussian fits for the polystyrene (blue circles) and silica particles (orange squares), with error-bars showing the standard deviations. Theory predicts the dashed curves, which agree very well with the measurements, without any free parameters. The calculated values of $\rho$ and $\Sigma\tau_{0m}$ are plotted in Methods (Figure S1).

The time-of-flight for particles travelling through the fibre depends on the strength of the optical scattering force $F_s$, which can be calculated from the Poynting vector difference between the two sides of the particle[22]. The analytically calculated $F_s$ agrees well with a three-dimensional finite element simulation for the Maxwell stress tensor (see Methods, Figure S2). Once inside the fibre core, the particle (velocity $v_p$) accelerates exponentially towards a constant velocity $v_{pT} = F_s/\gamma$, at which point the optical and viscous forces balance:

$$v_p(t) = v_{pT}\left(1 - \exp(-\gamma t/m)\right) \quad (3)$$

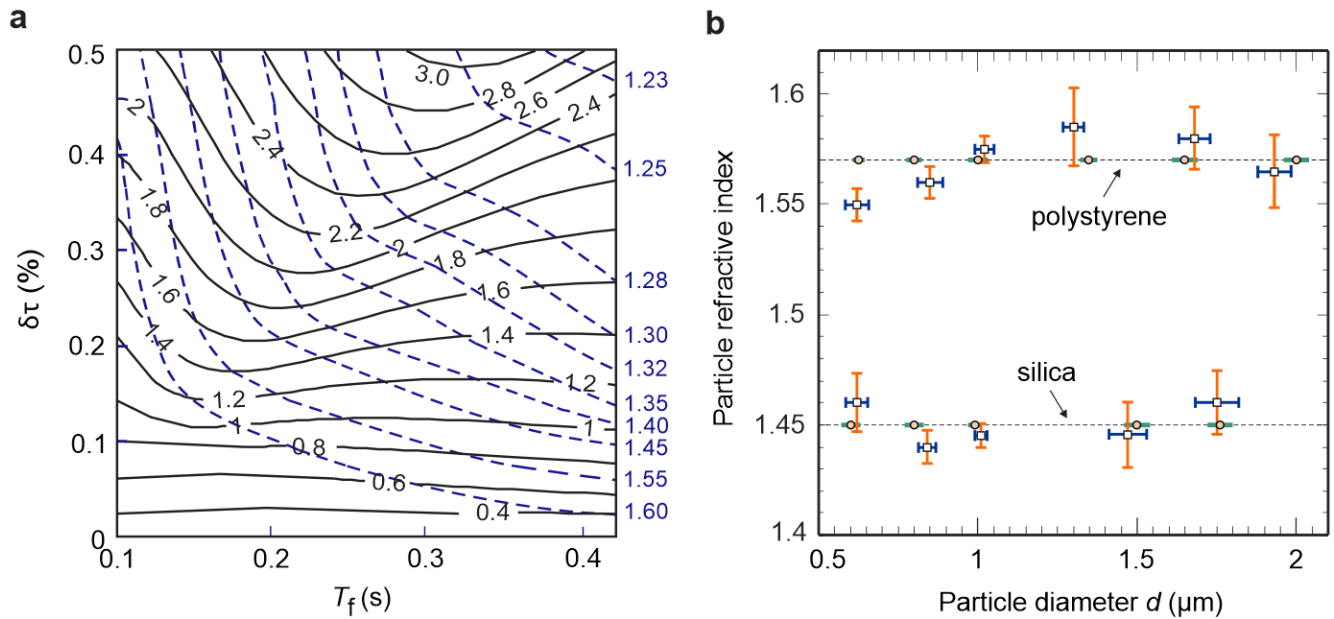

**Fig. 3 | Measurements of particle diameter $d$ and refractive index $n_P$. a,** Contour plots mapping the particle diameter $d$ (black full curves) and refractive index $n_p$ (dark blue dashed curves) to transmission drop $\delta\tau$ and time-of-flight $T_f$. Unambiguous estimates of both $d$ and $n_p$ are possible. **b,** The white squares mark the retrieved particle diameter and refractive index, and the horizontal (blue) and vertical (orange) bars mark the standard errors. The pink dots mark the particle diameter $d$ and refractive index, and the green error-bars the standard deviation of $d$, specified by the manufacturer.

where $m$ is the mass of the particle, $\gamma = 3\pi\eta d$ the drag coefficient and $\eta$ the air viscosity, corrected for the hollow core[23]. For typical experimental parameters (fibre length $L = 10$ cm, optical power 3 W, atmospheric pressure, a silica particle with $d = 1$ μm, hollow core diameter D = 20 μm, $\eta = 17.9$ μPa.s), the acceleration time $m/\gamma$ is $8.7\times10^4$ times shorter than $T_f$, which is therefore given to a very good approximation by $\gamma L/F_s$. The lower plot in Figure 2b compares theoretical (dashed lines) with measured values of $T_f$ for polystyrene (blue circles) and silica particles (orange squares), showing excellent agreement.

Figure 3a shows that particles with different combinations of refractive index np (dark blue dashed contours) and diameter d (black contours) are conveniently located at different regions on the $\delta\tau$ versus Tf map. Theory can therefore be used to construct a look-up table that unambiguously translates the measured $\delta\tau$ and Tf into d and np. The white squares in Figure 3b show the retrieved particle diameter and refractive index for the polystyrene and silica particles, using the measured mean values of $\delta\tau$ and Tf from the histogram. The pink dots mark the refractive indices and diameters specified by the manufacturer. The error-bars (blue and orange) mark the estimated measurement precision σd in particle diameter and σn refractive index, as explained in Methods. For a 0.99 μm silica particle, the measurement errors are as small as ±0.9% for the diameter and ±0.2% for the refractive index.

In conclusion, real-time counting, sizing and refractive index measurement of airborne PM2.5 particles can be carried out with high accuracy in HC-PCF. The unique ability to measure the refractive index will assist in identifying the particle material. Fitted with a suitable particulate air filter at the input to remove larger particles, the system offers a compact and cost-effective means of continuously monitoring air-borne PM2.5 particles in many real-world contexts, e.g., urban areas and industrial sites. Since the optical trapping prevents particles from adhering to the core wall, the system also offers very long (perhaps unlimited) operating lifetimes, in addition the particle detection rate could be increased by using an array of HC-PCFs. Finally, the technique is also suitable for operation in aqueous environments[24], with applications for example in monitoring water pollution.

# Methods

## 1. Experimental configuration

Light from a 3W CW laser at 1064 nm wavelength was free-space coupled into the $LP_{01}$ core mode of the HC-PCF to trap and propel airborne particles through the fibre. The fibre transmission was monitored by a photodetector (PD) after attenuation and spatial filtering using an aperture. Fig. 1a shows the measured near-field intensity profile of the core mode superimposed on a scanning electron micrograph of the kagomé-style HC-PCF (core diameter 17.7 µm). The fibre length used in the experiments was 7.3 cm.

The process of loading particles was similar to that reported in [22]. At first the particles were added to distilled water in a particle-to-water mass ratio of 0.001 to 0.01. To avoid particle clustering (caused by van der Waals forces) and improve the homogeneity, the suspension was then placed in an ultrasonic bath for 30 mins at 40ºC. A medical nebulizer (OMRON NE-U22-E) with a mesh grid size of 4.2 µm was used to produce droplets from the particle suspension, the concentration being chosen so that each droplet contained on average a single particle. Even after this treatment clusters did sometimes form, especially for smaller particles, and if they were less than 4.2 µm in diameter they could pass through the mesh and be trapped. The clusters detected in Figure 2a are thus mainly aggregates of smaller particles. Particles were injected via an inlet tube on the glass lid, which was positioned above and aligned with the fibre endface. In this way, the loaded particles have a high probability of entering the beam and being trapped. Once a droplet is captured, laser heating very rapidly vaporizes the water, resulting in a dry trapped particle.

## 2. Forward and backward contributions to transmission drop

Figure S1 shows the calculated contributions to $\delta\tau$ from forward intermodal (green) and backward (orange) scattering for polystyrene particles, plotted against particle diameter. The sum (blue-dashed) agrees well with the measurements.

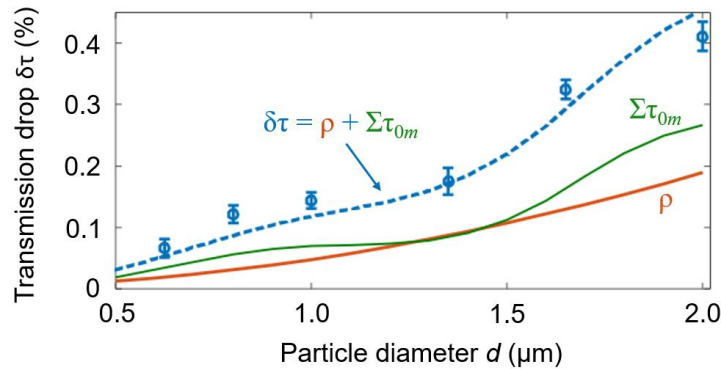

**Fig. S1.** Simulated reflected ($\rho$) and forward intermodal ($\Sigma\tau_{0m}$) contributions to the transmission drop. The sum of those two, $\delta\tau$, matches very well to the experimental data-points.

## 3. Optical force calculation

Figure S2 compares analytical values [22] for the optical force $F_s$ acting on polystyrene particles with the results of 3D finite element modelling (FEM) using the Maxwell stress tensor. The agreement is excellent. The fluctuations in $F_s$ predicted by FEM are caused by Mie resonances in the particles.

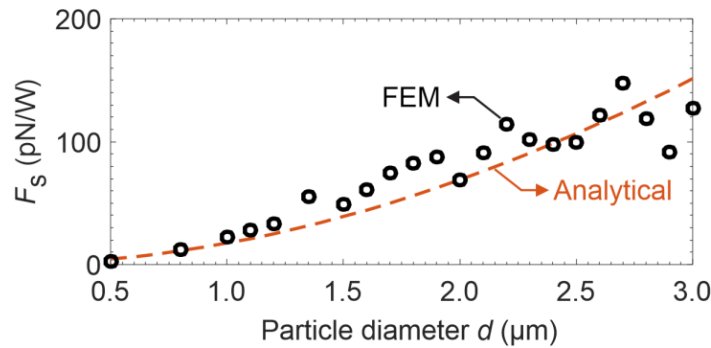

**Fig. S2.** Comparison between analytical (orange-dashed line) and numerical (FEM, black circles) values of the optical scattering forces for polystyrene particles, plotted as a function of $d$. The hollow core diameter is 17.7 µm, the wavelength is 1064 nm and the incident light is in the $LP_{01}$ mode.

## 4. Histogram of transmission drop and time-of-flight for silica particles

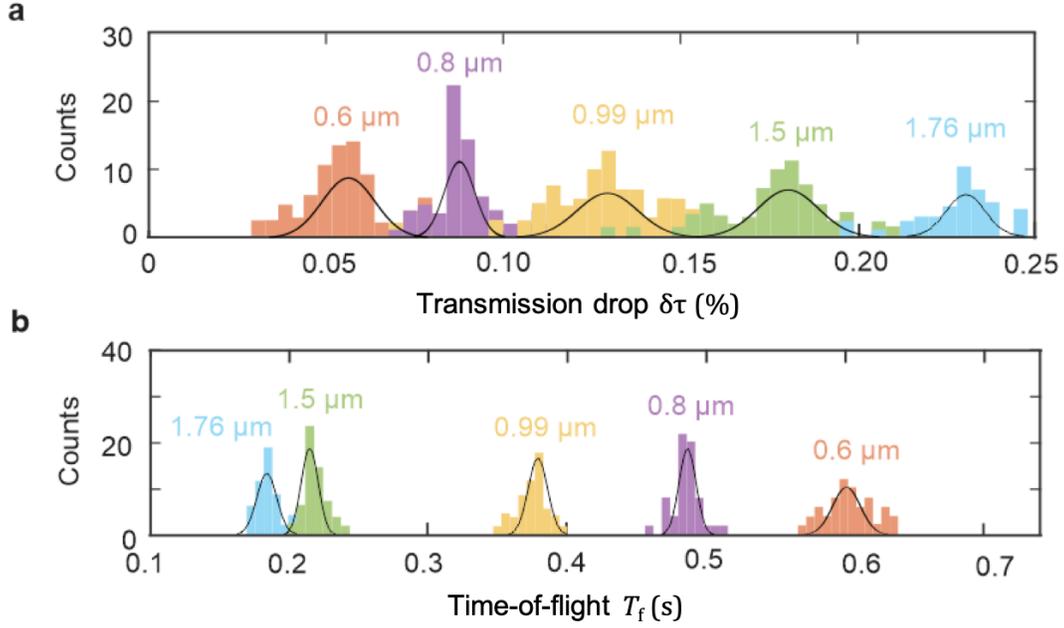

**Fig. S3.** Measured histograms of (a) the transmission drop and (b) time-of-flight for silica particles. The black curves represent fits to Gaussian distributions.

## 5. Estimate of measurement precision of particle size and refractive index

The standard deviations of the retrieved values of $d$ and $n_p$ ($\sigma_d$ and $\sigma_n$) were estimated using:

$$\sigma_d = \sqrt{\left(\frac{\partial d}{\partial (\delta\tau)}\right)^2 \sigma_{\delta\tau}^2 + \left(\frac{\partial d}{\partial (T_f)}\right)^2 \sigma_{Tf}^2}$$

$$\sigma_n = \sqrt{\left(\frac{\partial n_p}{\partial (\delta\tau)}\right)^2 \sigma_{\delta\tau}^2 + \left(\frac{\partial n_p}{\partial (T_f)}\right)^2 \sigma_{Tf}^2}$$

(S1)

where $\sigma_{\delta\tau}$ and $\sigma_{Tf}$ are the measured standard deviations of $\delta\tau$ and $T_f$. The partial derivatives were obtained by numerically calculating the two-dimensional gradient of $d$ and $n_p$ with respect to $\delta\tau$ and $T_f$ for each individual data-point in the contour map shown in Figure 3a.